\documentclass[aps,prx,twocolumn,a4paper,floatfix,showpacs]{revtex4}
\usepackage{amsmath}
\usepackage{graphicx}
\usepackage{amsfonts}
\usepackage[utf8]{inputenc}
\usepackage{color}
\usepackage{cancel}
\usepackage{hyperref}
\graphicspath{{fig/}}

\begin{document}

\title{Statistical mechanics of multi-edge networks}

\author{O. Sagarra}
\author{C. J. Perez-Vicente}
\author{A.  D\'{\i}az-Guilera}
\affiliation{Departament de F\'{\i}sica Fonamental, Universitat de Barcelona, 08028 Barcelona, Spain}

\begin{abstract}
Statistical properties of binary complex networks are well understood and recently many attempts have been made to extend this knowledge to weighted ones. There is, however, a subtle difference between networks where weights are continuos variables and those where they account for discrete, distinguishable events, which we call multi-edge networks. In this work we face this problem introducing multi-edge networks as graphs where multiple (distinguishable) connections between nodes are considered. We develop a statistical mechanics framework where it is possible to get information about the most relevant observables given a large spectrum of linear and nonlinear constraints including those depending both on the number of multi-edges per link and their binary projection. The latter case is particularly interesting as we show that binary projections can be understood from multi-edge processes. The implications of these results are important as many real agent based problems mapped onto graphs require of this treatment for a proper characterization of its collective behavior. 
\end{abstract}
\pacs{a}
\maketitle

\section{Introduction:}
\label{sec_intro}

The increasing and unprecedented quality and quantity of available data coming from very different areas is boosting the field of complex networks. Interdisciplinar science demands new efforts and new tools, addressed not only to develop more efficient computational strategies to analyze incoming data but also to span a theoretical framework where both more accurate and more tractable models can provide predictions closer to the reality one wants to face.
In this context, a standard approach consists in representing in a graph the complex structure of interactions among the elements of a given system. Statistical mechanics is an extraordinary framework where such a complex structure can be appropriately modelled and with this aim a large amount of studies have appeared in the last years \cite{Albert2007,Newman2010,Dorogovtsev2003}.

The simplest representation of a network assumes the existence of nodes and edges. The edges are not necessary symmetric, they can give us information about the relative influence (interaction) of a node onto another, i.e., they can be directed and have a certain strength. However, the first studies where essentially focused on a binary projection of the network on a graph where only the existence of an edge and its distribution where required to determine some properties of the network, being weights out of consideration. In this way one could compute the probability that two arbitrary sites (nodes) were connected through an edge.
It is well known that such probability keeps certain analogies with occupation numbers in quantum statistics, in particular with \textit{fermionic} systems. Further developments have extended these results to richer and more complex structures such as directed and weighted graphs finding analogy with \textit{bosonic} systems.



Yet a fundamental discussion remains to be done about the nature of the entities forming the systems considered. A successful and complete description of any system susceptible to be represented as a complex network through statistical physics requires an appropriate characterization of the features of the microscopic components at hand. In this context it is fundamental to know whether simple units can be perfectly identified for the a proper creation of a graph in analogy to statistical mechanics where the distinguishability of particles leads to different descriptions in terms of Fermi-Dirac, Bose-Einstein or Maxwell-Boltzmann statistics. An example where the characterization of individuals is crucial is the transportation network \cite{Barthelemy2011,Kaluza2010,Roth2011}. Processes generated by single agents represent single events, for instance a trip between two locations or in social sciences one could also talk about an event as a call or an email, and allow to build the so-called \textit{Origin Destination} matrices which collect global information about their mobility or allocation. A naive approach based on a standard weighted description \cite{Park2004,Garlaschelli2009} of a network is not satisfactory for this problem as it was pointed out by Wilson \cite{Press1967} who mapped transport systems to statistical physics using an entropy maximization approach.

\begin{figure}[hbp]
\begin{center}
\includegraphics[scale=0.3]{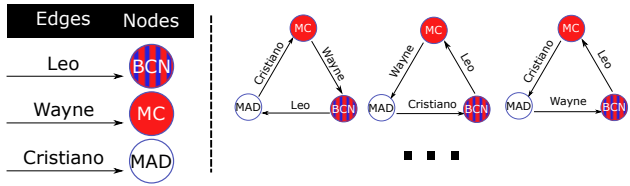}
\caption{Example of configurations in a given multi-edge network with distinguishable edges: Out of a set of three distinguishable nodes and three distinguishable (directed) events, one can see that the number of available micro-states is higher than in the usual case, since all permutations of events generate \textit{different} networks.}
\label{fig1}
\end{center}
\end{figure}

The present work addresses this issue by presenting a statistical mechanics approach of networks created from distinguishable single-unit events that can be grouped in edges. The difference between this representation, which leads to what we call multi-edge networks, and the already considered \textit{weighted} ones is clear as schematically shown in figure \ref{fig1}: Multi-edge networks assume the existence a minimal \textit{weight} of unity value (and hence a quantization) representing an indissociable \textit{event} of distinguishable and independent nature; in some cases, groups of these events connect the same pair of nodes, hence forming an edge with multiple connections, which have different nature from a \textit{weighted} edge where neither quantization is imposed nor any obligatory distinguishability of entities forming it. The choice of one or the other representation will thus depend on the problem at hand and makes a big difference in terms of collective behavior of the whole network.

The paper is structured as follows: Section \ref{sec_nomenclature} introduces the concept of multi-edge networks together with some basic nomenclature and definitions used. Sections \ref{sec_cano}, \ref{sec_grandca} and \ref{seq_kss} introduce the different methodologies used to solve the micro-canonical ensemble under different sets of constraints developing at each time a different example to work on (mainly linear constraints on occupation numbers, binary-projection constraints and both). Finally some concluding remarks are given discussing the work done. The appendices contain details, additional discussion and some mathematical developments on mentioned examples in the text.

\section{Ensemble approach to multi-edge networks: Nomenclature and definitions}
\label{sec_nomenclature}
A multi-edge network is a collection of nodes that may be connected by none, one or more than one element (\textit{event}) taken from a set of independent, distinguishable entities.  Such an object admits a coarse-grained representation in terms of a multi-edge matrix \textbf{T}. The matrix entries $t_{ij}$ are bounded integer stochastic variables $t_{ij}\leq t_{ij}^{(\text{max})}=T$ denoting the number of \textit{events} joining nodes $i$ and $j$ and $T=\sum \limits_{ij} t_{ij}$ represents the total number of events. In the context of statistical mechanics they play an analogue role to \textit{occupation numbers}. The matrix $\mathbf{T}$ need not necessarily be symmetric so $t_{ij} \neq t_{ji}$ in general. Let $N$ be the number of nodes of the network which \textit{generate} \footnote{Please note that the framework we present permits self-edges if one desires to, it is just a matter of extending the sums over available states over all values of $i,j$ or exclude the term $i\neq j$. It also is valid for both directed and undirected networks: One needs to perform such sums only over states for which $j<i$ in such cases.} $N(N-1)$ possible states where individual events can be allocated. $E=\sum \limits_{i\neq j} \Theta(t_{ij})$ represents the number of \textit{occupied} states (regardless of their occupation number as long as $t_{ij}>0$) in the network, representing $\Theta(x)$ the Heavyside step function
\begin{equation}\nonumber
\Theta(x) = \left \{ \begin{array}{l l} 0 & x=0 \\ 1 & x\geq 1  \end{array} \right. \quad (x\in \mathcal{N}).
\end{equation}
Such function ensures that $E$ accounts for the number of existing connections regardless of the number of events contained in each entry of the matrix, therefore creating a binary projection of the network.


Our goal is to construct an ensemble framework that allows to treat multi-edge networks with any given set of constraints $\vec C = \vec C(\{t_{ij}\})$ defined in terms of the variables of the system. The constraints define a macro-state and restrict the available phase space to all the possible graphs compatible with such constraints. In this context, we assume two main starting hypothesis. First, all the configurations (micro-states) compatible with the observed constraints have the same \textit{a priori} probability of appearance. Secondly, we assume that $T,N >> 1$ being $N$ given (this means that the topological structure of the network, the number of available states, does not change) which allows a statistical treatment of the problem. We further assume that the distribution of occupation numbers is stationary and defines in turn a probability that fixes our thermodynamic limit. This probability $p(t_{ij})$ indicates the asymptotic (relative) distribution of occupation numbers:
\begin{equation}\label{eq_thermo_lim}
p^*_{ij}= \left \langle \frac{t_{ij}}{ \sum t_{ij} } \right \rangle \to p(t_{ij}) \text{ as }T \to \infty.
\end{equation}

Where $\langle ... \rangle$ denotes an average performed over the ensemble considered.
Under these considerations it is possible to establish a complete mapping between the problem at hand and a classical statistical mechanics problem. The \textit{events} correspond to distinguishable particles occupying any of the $N(N-1)$ available energy states (the methodology is laid for directed graphs, but can be easily adapted to undirected ones, yielding $N(N-1)/2$ available states). The main difference between the proposed system and the ones studied by classical statistical equilibrium mechanics regards the constraints used: They will not be in general of extensive, global nature but local at level of nodes. The standard procedure, already used by different authors, consists in finding an expression for the probability of obtaining a macro-state defined by its multi-edge adjacency matrix $\mathbf{T}$,
\begin{equation}\nonumber
\mathcal{P}(\mathbf{T})=\mathcal{P}(\{t_{ij}\})
\end{equation}
and then maximize its associated entropy,
\begin{equation}\label{eq2}
S_{SH} = - \sum \limits_{\Gamma}\mathcal{P}(\mathbf{T}) \ln \mathcal{P}(\mathbf{T})
\end{equation}
where the sum runs over all possible configurations of \textit{events} (micro-states) on the accessible phase space $\Gamma$ of the given ensemble. Since distinguishability plays an important role in our work, it is important to note that if one wishes to compute the sum over possible values of the occupation numbers $\{t_{ij}\}$ (our canonical variables) rather than over configurations, a degeneracy term needs to be added to expression \eqref{eq2},
\begin{equation}\label{eq4}
\mathcal{D}(\{t_{ij}\}) = \frac{\left(\sum \limits_{ij} t_{ij}\right)!}{\prod \limits_{ij} t_{ij}!}.
\end{equation}
In the remainder of the paper, we shall work on the occupation-number space $\Omega$ (which is a coarsened representation of the $\Gamma$ configurational space), including the degeneracy term and hence considering an expression for the entropy of the form,
\begin{equation}\label{eq_entropy_tij}
\begin{split}
S_{SH} &= - \sum \limits_{\Gamma} \mathcal{P}(\{t_{ij}\}) \ln \mathcal{P}(\{t_{ij}\})= \\
&= - \sum \limits_{\Omega(\{t_{ij}\}} \mathcal{D}(\{t_{ij}\}) \mathcal{P}(\{t_{ij}\}) \ln \big{(}\mathcal{D}(\{t_{ij}\}) \mathcal{P}(\{t_{ij}\}) \big{)} =\\
&=- \sum \limits_{\Omega(\{t_{ij}\})} P(\{t_{ij}\}) \ln P(\{t_{ij}\})
\end{split}
\end{equation}
where we changed the notation $\mathcal{P} \to P$ to denote the counting over occupation numbers in the (coarse-grained) $\Omega$ space rather than event-configurations in $\Gamma$ space.

A final comment deserves the seminal work by Wilson on transport theory \cite{Wilson2010}. He followed a similar scheme and found expressions for the expected number of events connecting two arbitrary nodes according to some considered constraints. However, in the present paper we go further, we present a modern methodology able to handle more complicated constraints as compared to those analyzed by Wilson and also consider constraints which affect simultaneously the distribution of occupation numbers and its binary projection on the graph, for instance those affecting the strength (\textit{weighted} degree \cite{Barthelemy2005a}) of a node and its degree.


\section{Multi-edge network with given linear constraints on the occupation numbers $t_{ij}$}
\label{sec_cano}

In this work we proceed following a micro-canonical scenario. In this ensemble, all the configurations are equally probable and the constraints are considered "hard", i.e. the phase space accessible always fulfils strictly the constraints. Thus, we must only maximize the expression $\ln \Omega(\{ t_{ij} \})$ with respect to $t_{ij}$ which allows to find the most probable value of some relevant observables, i.e. their statistically expected value in the ensemble of maximally random graph which \textit{strictly} fulfil the constraints.

\begin{equation}\label{eq5}
\text{max} \left \{ \Omega'( \{ t_{ij} \} ) | \vec C ( \{ t_{ij} \} ) = \vec C \right \} \equiv \text{max}\left \{ \Omega( \{t_{ij} \}) \right \}.
\end{equation}
We follow the same methodology developped in seminal works by Bianconi \cite{Bianconi2008} and write down the volume of the $\Omega$ space introducing auxiliary fields $h_{ij}$.
\begin{equation}\label{eq7}
\begin{split}
\Omega& = \sum \limits_{\{c\}} \prod\limits_{q}^Q \delta(C_q - C_q(\{t_{ij}\})  e^{\sum\limits_{\{i,j\}} h_{ij}t_{ij}}=\\
& = \sum \limits_{\{c\}} \prod\limits_{q}^Q \int d\vec\theta_q e^{\theta_q( C_q(\{t_{ij}\}-C_q)}e^{\sum\limits_{\{i,j\}} h_{ij}t_{ij}}
\end{split}
\end{equation}
where ${\{c\}}$ denotes a sum over microscopic configurations ($\Gamma$ space), $Q$ is the total number of constraints and $\vec\theta_q$ are the related Lagrange multipliers. Note that also an integral representation of the Kronecker delta has been used. The introduction of auxiliary fields $h_{ij}$ allows to recover all the central moments of the distribution of $t_{ij}$. 
In fact, one can see that \eqref{eq7} is closely related to the cumulant generating function of $t_{ij}$ and hence all its cumulants can be recovered by differentiation. In particular,
\begin{equation}
\begin{split}\label{eq_der_cum}
\langle t_{ij} \rangle &= \left . \partial_{h_{ij}} \ln\Omega(\{t_{ij}\}) \right |_{h_{ij}=0 \, \forall i,j} \\
\sigma^2_{t_{ij}} &= \left .\partial^2_{h_{ij}} \ln\Omega(\{t_{ij}\}) \right |_{h_{ij}=0 \, \forall i,j}\\
\sigma^2_{t_{ij},t_{kl}} &= \left . \partial^2_{h_{ij}h_{kl}} \ln\Omega(\{t_{ij}\}) \right |_{h_{ij}=0 \, \forall i,j} .
\end{split}
\end{equation}

If we wish to perform the sum over occupation numbers ($\Omega$ space) rather than over configurations of the system, we need to take into account the degeneration given in \eqref{eq4},
\begin{equation}
\begin{split}\label{eq8}
\Omega & = \sum \limits_{\{t_{ij}'\}} \prod\limits_{q}  \int d\vec\theta_q e^{-\theta_q C_q} \frac{T!}{\prod\limits_{ij} t'_{ij}!} e^{\theta_q C_q(\{t'_{ij}\})}e^{\sum\limits_{ij} h_{ij}t'_{ij}}.
\end{split}
\end{equation}
For this ensemble and for linear constraints on occupation numbers, one can write
\begin{equation}\label{eq_cons_cann}
C_q(\{t_{ij}\})= \sum_{ij} c^{(ij)}_q t_{ij}
\end{equation}
being $c_q^{(ij)}$ a quantity that usually depends on a "property" of the edge between nodes $i$ and $j$ (a distance for instance) or be a real number ($c^{(ij)}_q = \delta_{i,q}$ for out-going strength sequence constraints for example, as we shall see). In such situation, the sum over $\{t_{ij}\}$ sequences such that $\sum t_{ij}=T$ can be exactly performed inside the integral in \eqref{eq8} yielding,
\begin{equation}\label{eq_cumul_multi}
\begin{split}
\Omega & = \int \left ( \prod\limits_{q}  d\vec\theta_q\right ) e^{-\sum\limits_q \theta_q C_q} e^{T \ln \sum\limits_{ij} \exp \left\{(h_{ij}+\sum \limits_q \theta_q c^{(ij)}_q)\right\}}= \\
&=\int \left ( \prod\limits_{q}  d\vec\theta_q\right ) e^{f(\{\theta_q, C_q\}, \{h_{ij}\})} .
\end{split}
\end{equation}

The occupation number statistics can be shown to have multinomial nature (see appendix \ref{ap_cumul}),
\begin{equation}
\begin{split}\label{eq_multi_prop}
\langle t_{ij} \rangle &=  Tp_{ij}; \quad \\
\sigma^2_{t_{ij}} &= Tp_{ij} \left ( 1 - p_{ij} \right ) \\
\sigma^2_{t_{ij},t_{kl}} &= - T p_{ij}p_{kl}
\end{split}
\end{equation}
This fact assures that the relative fluctuations of occupation numbers vanish in the thermodynamic limit ($T\to \infty$) since,
\begin{equation}\nonumber
\frac{\sigma_{t_{ij}}}{\langle t_{ij}\rangle} = \sqrt{\frac{1-p_{ij}}{p_{ij}T}} \to 0 \quad \quad \text{as } T\to \infty, 
\end{equation}
We identified $p_{ij}$ as the probability for an individual event to be assigned to state $ij$. Explicitly,
\begin{equation}\label{eq_pij_multi}
p_{ij}=  \frac{e^{\sum\limits_q c_q^{ij} \theta_q}}{\sum\limits_{ij} p_{ij}}.
\end{equation}


Concerning the entropy of the graphs in this ensemble, the integral in \eqref{eq_cumul_multi} can be approximated to first order by using steepest descent methods,
\begin{equation}\label{eq_entropy_can}
\begin{split}
S_{BG} &= \ln \left. \Omega\right|_{h_{ij\, \forall i,j}=0} \simeq - f^* = \\
&=-\sum\limits_{q} C_q \theta^*_q + T \ln e^{\sum_q \theta^*_q \sum \limits_{ij} c_q^{(ij)}}
\end{split}
\end{equation}
where $\theta^*_q$ are the solutions of the saddle point equations \footnote{In the remainder of the paper, the $*$ signs will be omitted to simplify notation.} given by,
\begin{equation}\label{eq_max_const}
\begin{split}
\left. \partial_{\theta_q} f\right|_{h_{ij}=0 \, \forall ij} &= 0 \implies C_q = C_q(\langle t_{ij} \rangle) = T \sum_{i,j} c_q^{(ij)} p_{ij}.
%
\end{split}
\end{equation}

Merging \eqref{eq_multi_prop}, \eqref{eq_pij_multi}, \eqref{eq_entropy_can} and \eqref{eq_max_const} one obtains the event-specific entropy of a given graph in this ensemble,
\begin{equation}\label{eq_entro_can}
\begin{split}
\frac{S_{BG}}{T} &= - \left ( \sum_q \theta_q \sum_{i,j} c_q^{(ij)} p_{ij}  - \ln \sum \limits_{ij} e^{\sum_q \theta_q c_q^{(ij)}}  \right)= \\
&=- \left ( \sum_{ij} p_{ij} \sum_q \theta_q c_q^{(ij)}   - \ln \sum \limits_{ij} e^{\sum_q \theta_q c_q^{(ij)}}  \right) =\\
&= - \sum \limits_{ij} p_{ij} \ln p_{ij} = \frac{S_{SH}}{T}.
\end{split}
\end{equation}
which has a final Shannon entropy form, equivalent to the Boltzmann-Gibbs entropy.

If one is able to exactly solve the saddle point equations obtained from the steepest descent approximation, 
the full distribution of occupation numbers is recovered. Despite being in a micro-canonical framework, one could consider a \textit{canonical} ensemble where the constraints are fulfilled on average, hence $C_q(\langle t_{ij} \rangle) = \langle C_q (t_{ij}) \rangle$. Having proven that the partition function \eqref{eq_cumul_multi} has a multinomial-cumulant form over the occupation numbers and considering only linear \textit{soft} constraints, the requirement that those constraints need to be fulfilled only on average in the sampling over the phase space is automatically satisfied. Moreover, the constraints have vanishing relative fluctuations in the thermodynamic limit,
\begin{equation} \nonumber
\begin{split}
\langle C_q(\{t_{ij}\}) \rangle&= \sum_{i,j} c^{(ij)}_q \langle t_{ij} \rangle = T \sum_{i,j} c^{(ij)}_q p_{ij} \\
\sigma^2_{C_q}&= \sum_{i,j,k,l} c^{(ij)}_q c^{(k,l)}_q \sigma^2_{ij,kl}  \propto T\\
\frac{\sigma^2_{C_q}}{\langle C_q \rangle^2} &\to \frac{1}{T} \to 0 \text{ as }T\to \infty,
\end{split}
\end{equation}
where we have used the properties of the multinomial distribution presented in \eqref{eq_multi_prop}. Although the theoretical basis for the generation of graphs in different ensembles is introduced in this paper (see appendix \ref{ap_equiv}), the challenges for the exact and efficient generation of such ensembles will be shortly tackled and presented in future work.

In the following, we shall consider some explicit cases of linear constraints on $t_{ij}$.
\subsection{No constraints}
\label{fixT_cano}
This is the simplest case where we have a single hard constraint (apart from the number of nodes $N$) $T=\sum\limits_{ij} t_{ij}$. Therefore \eqref{eq8} reads,
\begin{equation}\nonumber
\begin{split}
\Omega=& \int d\vec \theta e^{-\theta T} \sum \limits_{\sum t_{ij}=T} \frac{T!}{\prod\limits_{ij} t_{ij}!} \prod\limits_{ij} \left (e^{\theta+h_{ij}}\right)^{t_{ij}} = \left (\sum \limits_{ij} e^{h_{ij}} \right )^T,
\end{split}
\end{equation}
where we have taken profit of the normalized structure of the integral. In this case it is straightforward to determine the average value of the occupation numbers
\begin{equation}\label{eq_nocons}
\begin{split}
\langle t_{ij} \rangle &= \frac{T}{(N(N-1))} \equiv \bar{t} \equiv Tp \quad \forall\, i,j.
\end{split}
\end{equation}
Therefore, all the occupation numbers have constant probability $p_{ij}=p \, \forall\,i,j$ of being \textit{chosen} per event sorted. The result is according to what intuition would tell us: Under no-constrains events are equally distributed among levels which reminds the high temperature regime in classical systems where there is an arbitrary large (but finite) number of energy levels. It is also possible to compute the covariances on occupation numbers,
\begin{equation}
\begin{split}\nonumber
&\sigma^2_{t_{ij},t_{kl}} = \left \{ \begin{array} {l l}  - T p^2 & ij\neq kl\\
 T p (1-p) & ij=kl \end{array} \right. .
\end{split}
\end{equation}
Other typical network magnitudes of interest such as the strengths (both incoming $s^{(in)}_j = \sum\limits_i t_{ij}$ and outgoing $s^{(out)}_i = \sum\limits_j t_{ij}$) , which will be extensively used in this paper, can also be computed. They are random integer variables with fixed mean $\langle s_i \rangle \equiv \bar{s} = (N-1)pT \equiv p_s T$ and variance $\sigma^2_{s} = Tp_s(1-p_s)$ $\forall i\in [1,N]$ which are \textit{on average} equal for each node and have also a multinomial character.  In this case $\bar{s}$ denotes an average over a single graph realization.

Concerning the Boltzmann-Gibbs entropy of the ensemble
\begin{equation}\label{eq_nocons_entr}
\begin{split}
S_{BG}&=  T \ln (N(N-1)) = - T\ln p =S_{SH}.
\end{split}
\end{equation}
Which recovers a Shannon form over events as expected, $S_{SH} = \sum\limits_{ji} p \ln p$.


\subsection{Fixed average event cost $\bar{c}=C/T$}
\label{fixTC}
If additionally to the number of events $T$, we consider a cost matrix (symmetric, dense and positive definite) $\mathbf{D}=\{d_{ij}\}$  and fix the total cost $C_T=\sum d_{ij} t_{ij}$, we trivially get,
\begin{equation}\nonumber
\begin{split}
\Omega&= \int d\vec\theta  \exp \left \{-\theta C_T + T \ln \sum e^{h_{ij}+\theta d_{ij}} \right \}=\\
&=\int d\vec\theta \exp \left \{ c(\theta,\{h_{ij}\},\{t_{ij}\})\right\},
\end{split}
\end{equation}
which leads to the saddle point equation
\begin{equation}\nonumber
\partial_\theta c |_{h_{ij}=0\, \forall ij}= 0 \implies \frac{C_T}{T}\equiv \bar{c} = \sum \frac{d_{ij} e^{\theta d_{ij}}}{\sum e^{\theta d_{ij}}},
\end{equation}
and finally,
\begin{equation}\nonumber
\begin{split}
\langle t_{ij} \rangle &= T \frac{e^{\theta d_{ij}}}{\sum e^{\theta d_{ij}}}= T p(\theta, d_{ij})\\
\sigma^2_{t_{ij}} &=  Tp( \theta d_{ij}) (1- p(\theta, d_{ij})).
\end{split}
\end{equation}
And this leads to a \textit{weighted} version of the Waxman graph \cite{12889}. 
This reasoning can be extended to study the interesting case where the distribution of costs is also fixed \footnote{As done in \cite{Bianconi2009} for the binary case.}, which is of particular interest in the field of \textit{O-D} matrices used to analyze mobility, where the mobility of users using certain types of transports is assumed to follow particular statistical forms (\cite{Bazzani2010,Liang2012a}). This case is analyzed in detail and solved in appendix \ref{ap_extra}.

For any of the cases involving cost matrices, specially those related with distances, it is very important to remark that the allocation of occupation numbers is not independent in each state and hence it is not true that the probability $W(t_{ij},d_{ij},\theta)$ of having a state occupied by $t_{ij}$ events at distance $d_{ij}$ is $W(t_{ij},d_{ij})= K f(\theta, d_{ij})$. Rather $W(t_{ij})$ represents a conditional probability of observing $t_{ij}$ events (\textit{trips} in this scenario) at distance $d_{ij}$ given the distance matrix $\mathbf{D}$ and the rest of the constraints (included in $K$). This means in particular that if a \textit{deference} function $f(d_{ij})$ is proposed to explain observed flows between locations, as usually done in \textit{O-D} matrix studies under a maximum entropy assumption, its \textit{predictive} results need to be statistically tested against the full expected distribution of $t_{ij}$, and not only against the (biased) statistic of \textit{observed} or \textit{existing} occupation numbers.

\subsection{Fixed relative strength sequence $\vec s=\{(s^{out},s^{in})_i\}\,,T$}\label{fixS}
We consider now the case in which the only given constraint is the strength sequence $\vec s$. In this case, the constraints read,
\begin{equation}\label{eq_cons_ss}
C_i(t_{ij})= \sum_i t_{ij} = s_i^{out} \quad C_j(t_{ij})= \sum_j t_{ij} = s_j^{in},
\end{equation}
and equation \eqref{eq7} becomes,
\begin{widetext}
\begin{equation}\nonumber
\begin{split}
\Omega=& \int e^{-\sum_i \alpha_i s_i^{out} }e^{-\sum_j \beta_j s_j^{in} } \prod_{i} d\vec \alpha_i d\vec \beta_i
\sum \limits_{\sum t_{ij}=T} \frac{T!}{\prod\limits_{ij} t_{ij}!} \prod\limits_{ij} \left (e^{\alpha_i + \beta_j+h_{ij}}\right)^{t_{ij}} =\\
=&\int \prod_{i} d\vec \alpha_i d\vec \beta_i e^{-\alpha_i s_i^{out} }e^{-\beta_i s_i^{in} }
\left (\sum\limits_{ij} e^{\alpha_i+\beta_j+h_{ij}}\right)^T =\int d\vec \alpha d\vec \beta \exp \left (-\sum\limits_i\alpha_i s_i^{out}-\sum\limits_j \beta_j s_j^{in} \right.
+ \left. T \ln \left(\sum_{i\neq j} e^{\alpha_i+\beta_j+h_{ij}} \right) \right)=\\
=&\int d\vec \alpha d\vec\beta \exp f(\{\alpha_i\},\{\beta_i\},\{h_{ij}\}).
\end{split}
\end{equation}
\end{widetext}
Now we need to solve the $2N$ saddle point equations,
\begin{equation}\label{eq_sad_s}
\begin{split}
\left. \partial_{\alpha_i} f\right|_{h_{ij}=0 \, \forall i,j} &= 0 \implies s_i^{out} = T \sum_{j\neq i} \frac{e^{\alpha_i + \beta_j}}{\sum e^{\alpha_i+\beta_j}} \\
\left. \partial_{\beta_j} f \right|_{h_{ij}=0 \, \forall ij} &= 0 \implies s_j^{in} = T \sum_{i\neq j} \frac{e^{\alpha_i + \beta_j}}{\sum e^{\alpha_i+\beta_j}}
\end{split}
\end{equation}
And we apply again \eqref{eq_der_cum} to get,
\begin{equation}\label{unc_s}
\langle t_{ij} \rangle = T\frac{x_i y_j}{\sum\limits_{ij} x_i y_j} = T p_{ij},
\end{equation}
where we have identified $x_i= e^{\alpha_i}, y_j= e^{\beta_j}$.
and $p_{ij}$ has the same multinomial structure and properties as in previous examples except for the fact that these probabilities are state dependent. In fact, the average number of multi-edges between two nodes factorizes in an uncorrelated form.

For large $N$ we are led to,
\begin{equation}
\begin{split} \label{eq_fix_s}
s_i^{out} &= T x_i \frac{\sum\limits_j y_j}{\sum\limits_{i\neq j} x_i \sum\limits_j y_j } = T \frac{x_i}{X-x_i} \approx \frac{T x_i}{X}\\
X&\equiv \sum\limits_i x_i; \quad Y\equiv \sum\limits_j y_j; \quad s_i^{out} \propto x_i ; \quad s_j^{in} \propto y_j.
\end{split}
\end{equation}
and we recover the weighted configuration model \cite{serrano:101},
\begin{equation}\label{eq_we_conf}
\langle t_{ij} \rangle= \frac{s^{out}_i s^{in}_j}{T}.
\end{equation}
This result recovers the expression in \cite{Serrano2006} and is in accordance with a Maximum Likelihood Principle \cite{Garlaschelli2008a} (contrary to the case of \textit{weighted} networks as explained in \cite{Squartini2011b}).



Let us notice that these results allow a straightforward extension to the canonical ensemble. By identifying from \eqref{eq_fix_s}
\begin{equation}
p_{s^{out}_i} = \frac{x_i}{X} \quad \quad p_{s^{in}_j} =\frac{y_j}{Y},
\end{equation}
one can work with a strength sequence which is no longer fixed but a collection of fluctuating integer random values with a multinomial structure, since the partial grouping of multinomial random variables $t_{ij}$ preserve their multinomial character and by construction $\sum_j p_{s^{in}_j} = \sum_i p_{s^{out}_i} = 1$. Therefore,

\begin{equation}\label{eq_can_s_m}
\begin{split}
\langle s_i^{out} \rangle = Tp_{s^{out}_i} \quad & \quad \quad \langle s_j^{in} \rangle= T p_{s_j^{in}}\\
\sigma_{s_i^{out}} = Tp_{s^{out}_i}(1-p_{s^{out}_i}) \quad &\quad \quad \sigma_{s_j^{in}} = Tp_{s^{in}_j}(1-p_{s^{in}_j}). \\
\end{split}
\end{equation}

Concerning the entropy in this canonical ensemble scenario, making use of \eqref{eq_can_s_m} we further obtain a closed expression in terms of the constraints of the problem,
\begin{widetext}
\begin{equation}
\begin{split}
S_{BG} & \simeq -\sum\limits_{i} s^{out}_i \ln \frac{s^{out}_i}{T} - \sum \limits_j s_j^{in} \ln \frac{s_j^{in}}{T} +T \ln \sum \limits_{i\neq j} \frac{s_i^{out} s_j^{in}}{T^2} =-T\left ( \sum_i p_{s^{out}_i} \ln p_{s^{out}_i} + \sum_j p_{s^{in}_j} \ln p_{s^{in}_j}  \right)  \\
\frac{S_{BG}}{T} &=  -\sum_i p_{s^{out}_i} \ln p_{s^{out}_i} - \sum_j p_{s^{in}_j} \ln p_{s^{in}_j} = \frac{S_{SH}}{T}.
\end{split}
\end{equation}
\end{widetext}
Let us remind that in this context $p_{s_i}=\sum \limits_j p_{ij}$ represents the probability of a certain node $i$ to accumulate $\langle s_i \rangle$ (incoming or outgoing) events \textit{on average}. We therefore recover a Shannon form for the entropy in both micro-canonical and canonical ensembles, which scales with the total number of events $T$ and is node (but also state) specific (from equation \eqref{eq_entro_can}).


Further additional constraints can be added leading to different models. An example would be to merge the last two examples of a network living in a metric space (or any network where we can give a cost to the edges) with fixed "accessibility" of each node (\textit{distance-weighted strength}). We could also fix the average cost per trip $\bar{c}$ (global constraint) and the strength of each node (node local constraint). In this case one would obtain the \textit{popular} doubly constrained gravity model in several forms \cite{Erlander1990}, also Stouffers' intervening opportunities model and even the newly proposed gravity model \cite{Simini2012a} (by choosing an appropriate form for the cost function), see \cite{Press1967} for extended discussion). 

The case of the gravity law models deserves a closer look since an entropy maximization approach yields a form $\langle t_{ij} \rangle = x_i y_j f(\gamma,d_{ij})$ which is not equivalent (in general) to $\langle t_{ij}  \rangle = s^{\alpha} s^{\beta} f(\gamma,d_{ij})$ because the values of the multipliers depend on the particular spatial distribution of the considered nodes and their relative strengths. Hence, despite the success attained by these kinds of models to reproduce empirical data  \cite{Kaluza2010,Krings2009,Goh2012} an entropy maximization approach could unify the different sets of exponents observed in each study (see \cite{Barthelemy2011}).

\section{Multi-edge network with given nonlinear constraints on the binary projection of the occupation numbers $\Theta(t_{ij})$}
\label{sec_grandca}

Let us consider now more complex situations such as the case of non-linear constraints on $t_{ij}$. One of the most relevant objects to look at when dealing with complex networks is the degree distribution. It concerns only the existence of links between arbitrary nodes of the graph regardless the number of multi-edges between them. In the framework presented in this paper it can be worked out as a function of the binary projection of the occupation numbers on the graph so, in general, such constraints can be expressed as

\begin{equation}\label{eq_binary_cons}
\hat{C}_{q'} = \sum\limits_{ij} \hat{c}_{q'}^{(ij)} \Theta(t_{ij}).
\end{equation}

The main technical difficulty in dealing with these types of constraints is that they do not allow the summation in a multinomial form of the terms in $\{t_{ij}\}$ inside the integral of the partition function for exactly fixed $T$. A workaround to perform the summation can be found, however: Instead of making the multinomial sum at once, we proceed in two steps. We firstly introduce a Kronecker delta in integral form inside the integral of the partition function \eqref{eq7} with associated Lagrange multiplier $\theta$ which fixes the total number of events. We secondly allow the sum inside the integral to cover all the available values of the phase space ($\{t_{ij} \in [0,T] \,\forall \, i,j \}$. Finally, since the grouping of terms in the sum not fulfilling the constraints (including the constraint on the total expected number of events) will be penalized by the Kronecker deltas introduced earlier, we relax the limit on the sum over individual occupation number configurations from $T\to\infty$ what reminds the standard approach to the grandcanonical ensemble (in appendix \ref{ap_finite} the calculation with finite $T$ is also given).

Proceeding as explained we obtain a new version of equation \eqref{eq7}:
\begin{widetext}
\begin{equation}\label{eq_grandca}
\begin{split}
\Omega&= \int d\vec \theta  e^{-\theta T} e^{-\sum_{q'} \lambda_{q'} \hat{C}_{q'}} \prod_{q'}  d\vec \lambda_{q'} T!\sum\limits_{\{i,j\}'} \prod\limits_{ij}\frac{e^{\left(b_{ij} +\sum_{q'} \lambda_{q'} \hat{c}_{q'}^{(ij)}\right) \Theta(t_{ij})} e^{(h_{ij} + \theta)t_{ij}}}{t_{ij}!} =\\
&=\int d\vec \theta  e^{-\theta T} e^{-\sum_{q'} \lambda_{q'} \hat{C}_{q'}} \prod_{q'}  d\vec \lambda_{q'} T! \left [ e^{\left(b_{ij} +\sum_{q'} \lambda_{q'} \hat{c}_{q'}^{(ij)}\right)} \sum_{t_{ij}'=1}^T \frac{e^{(h_{ij}+\theta)t_{ij}'}}{t_{ij}'!} + \frac{1}{0!}\right ] \times\\
&\times\sum\limits_{\sum t_{ij} = T-t_{ij}'} \prod\frac{\left( e^{b_{ij} +\sum_{q'} \lambda_{q'} \hat{c}_{q'}^{(ij)}}\right)^{\Theta(t_{ij})}\left( e^{h_{ij}+\theta}\right)^{t_{ij}}}{t_{ij}!} =\\
&= \int d\vec \theta e^{-\theta T} e^{-\sum_{q'} \lambda_{q'} \hat{C}_{q'}} \prod_{q'} d\vec \lambda_{q'} T!\exp \left \{ \sum \limits_{ij} \ln \left \{ e^{b_{ij}+ \sum \limits_{q'} \lambda_{q'} \hat{c}_{q'}^{(ij)}} \left (e^{e^{h_{ij}+\theta}} -1\right)+1 \right \} \right\}.
\end{split}
\end{equation}
\end{widetext}
Where we have introduced auxiliary fields $b_{ij}$ for the binary projections of the occupation numbers: Both the complete set of cumulants of the occupation numbers and their binary projections can now be recovered by differentiation: For the occupation numbers using \eqref{eq_der_cum} and in the case of the binary projection,
\begin{equation}
\begin{split}\label{eq_der_cum_bin}
\langle \Theta(t_{ij}) \rangle &= \left . \frac{\partial}{\partial {b_{ij}}} \ln\Omega(\{t_{ij}\}) \right |_{h_{ij}=b_{ij}=0 \, \forall i,j} \\
\sigma^2_{\Theta(t_{ij})} &= \left . \frac{\partial^2}{\partial {b^2_{ij}}} \ln\Omega(\{t_{ij}\}) \right |_{h_{ij}=b_{ij}=0 \, \forall i,j}\\
\sigma^2_{\Theta(t_{ij}),\Theta(t_{kl})} &= \left . \frac{\partial^2}{\partial {b_{ij}b_{kl}}} \ln\Omega(\{t_{ij}\}) \right |_{h_{ij}=b_{ij}=0 \, \forall i,j} .
\end{split}
\end{equation}

From expression \eqref{eq_grandca} one can compute explicitly those values yielding,
\begin{equation}\label{eq_grandca_occ}
\begin{split}
\langle \Theta(t_{ij}) \rangle &\equiv \hat{p}_{ij} = \frac{e^{\sum_{q'} \lambda_{q'} \hat{c}_{q'}^{(ij)}}\left(e^{e^{\theta}}-1 \right)}{\Delta_{ij}}\\
\langle t_{ij} \rangle &= \frac{e^{e^\theta} e^\theta}{e^{e^\theta}-1} \hat{p}_{ij} \equiv \overline{t_{+}} \hat{p_{ij}}\\
\sigma^2_{\Theta(t_{ij})} &= \hat{p}_{ij} (1-\hat{p_{ij}}); \quad \sigma^2_{t_{ij}} = \langle t_{ij} \rangle (1+ e^\theta - \langle t_{ij} \rangle)\\
\sigma^2_{t_{ij},t_{kl}} &= \sigma^2_{\Theta(t_{ij}), \Theta(t_{kl})} = 0 \text{ if }ij\neq kl\\
\Delta_{ij} &\equiv e^{\sum_{q'} \lambda_{q'} \hat{c}_{q'}^{(ij)}}\left(e^{e^{\theta}}-1 \right)+1.
\end{split}
\end{equation}

For ease in notation, we perform the change $\rho \equiv e^\theta$ and we identify $\langle \Theta(t_{ij})\rangle$ with the probability $\hat{p}_{ij}$ of connection of nodes $i$ and $j$  \cite{Park2003} and $\overline{t_+}$ with the graph-average occupation of existing links. In appendix \ref{ap_cumul} we prove that the resulting partition function is equal to the cumulant generating function of the outcome of $N(N-1)$ independent \textit{zero-inflated Poisson processes} (ZIP \cite{Lambert1992}) with individual associated probability
\begin{equation}\label{eq_zip}
P(t|\hat{p},\rho) = (1-\hat{p})^{1-\Theta(t)} \left (\frac{\hat{p}}{e^\rho-1} \frac{\rho^t}{t!}\right)^{\Theta(t)},
\end{equation}
which in turn, regarding only the binary projection, corresponds to the outcome of independent Bernoulli processes with probabilities $\hat{p}$,
\begin{equation}
P(\Theta(t)|p) = \hat{p}^{\Theta(t)} (1-\hat{p})^{1-\Theta(t)}.
\end{equation}

From \eqref{eq_grandca_occ} one sees that in this case the multi-edge structure is completely determined by the binary constrained topology. Our coarse-grained description in terms of independent occupation numbers  implies that for each state, two outcomes can be considered: Either the edge does not exist (obviously with $0$ occupation) or it does exist, in which case the resulting (conditioned) statistics being Poisson with mean value $\langle t | t\geq 1 \rangle = \rho \frac{e^\rho}{e^\rho-1}$. 

\begin{figure}[hbp]
\begin{center}
\includegraphics[scale=0.45]{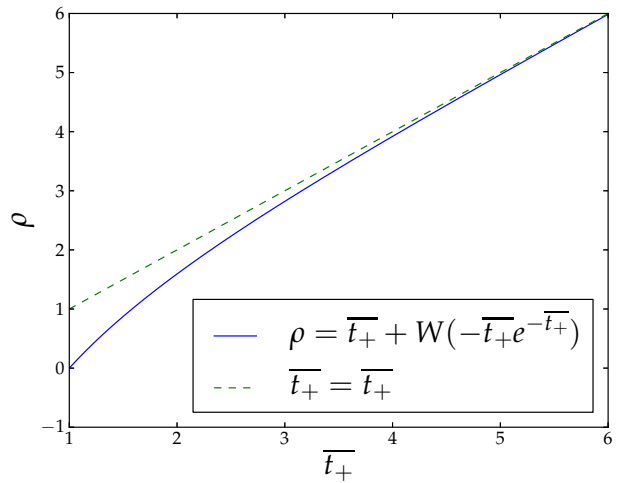}
\caption{The result of equation \eqref{eq_lambert} is shown together with the equality line $\rho = \overline{t_+}$. One can see the rapid convergence: For $\overline{t_+}=2.31$ we obtain $\frac{\rho}{\overline{t_+}}=0.9$ and for $\overline{t_+}=4.615$ one finds $\frac{\rho}{\overline{t_+}}=0.99$.}
\label{fig2}
\end{center}
\end{figure}

The constant relation of proportionality $\langle t_{ij} \rangle\propto \hat{p}_{ij}$ rapidly allows to identify the graph-average occupation of the expected existing links $\langle E \rangle = \sum\limits_{ij} \hat{p}_{ij}$,
\begin{equation}\label{eq_lambert1}
\begin{split}
&\overline{t_+} =  \frac{T}{\langle E \rangle} = \frac{e^\rho \rho}{e^\rho-1} > 1; \quad \rho> 0
\end{split}
\end{equation}
which can be inverted leading to,
\begin{equation}\label{eq_lambert}
\rho=W(-e^{-\overline{t_+}} \overline{t_+} ) + \overline{t_+}.
\end{equation}
Where $W(x)$ is the Lambert $W$ function \cite{Corless96onthe}. Figure \ref{fig2} shows a plot of equation \eqref{eq_lambert} stressing the rapid asymptotical convergence $\rho\to \overline{t_+}$ as $\overline{t_+}\to \infty$ (in fact, the approximation is clearly good as soon as $\overline{t_+}\simeq 5$).

With \eqref{eq_grandca_occ} and \eqref{eq_lambert} we can compute the relative fluctuations of both occupation numbers and binary links in the thermodynamic limit,
\begin{equation}\label{eq_rel_flu_mac}
\begin{split}
&\lim_{T\to \infty} \frac{\sigma^2_{\Theta(t_{ij})}}{\langle \Theta(t_{ij}) \rangle^2}  = \lim_{T\to \infty} \frac{(1-\hat{p_{ij}})}{\hat{p}_{ij}} = \frac{(1-\hat{p_{ij}})}{\hat{p}_{ij}}\\
&\lim_{T\to \infty} \frac{\sigma^2_{t_{ij}}}{\langle t_{ij} \rangle^2} = \lim_{T\to\infty} \frac{\left ( 1+\rho - \overline{t_+} \hat{p}_{ij}  \right )}{\overline{t_+} \hat{p}_{ij}} = \frac{1+\overline{t_+} (1-\hat{p}_{ij})}{\overline{t_+}\hat{p}_{ij}} \\ &\frac{\sigma^2_{t_{ij}}}{\langle t_{ij} \rangle^2} \to \frac{\sigma^2_{\Theta(t_{ij})}}{\langle \Theta(t_{ij}) \rangle^2} \quad \text{as } \hat{p}_{ij}\text{ fixed},\overline{t_+}=T/\sum\limits_{ij} \hat{p}_{ij} \to \infty.
\end{split}
\end{equation}

These expressions reflect the bimodal structure of the state statistics and explains the non-vanishing nature of the relative fluctuations: The variance of the occupation numbers has a maximum for $\hat{p}_{ij}|_{max} = \frac{1}{2} \left ( 1+ \overline{t_+}^{-1} \right) \to \frac{1}{2}$, vanishes for the absence ($\hat{p}\to 0$) of an edge and converges to Poisson statistics for edges that always exist ($\hat{p}\to 1$). The existence of an edge is a binary event, hence the maximum variability correspond to the draw situation ($50\%$ chance). In such a case, approximately half of the times a graph is created the considered edge will have (on average) occupation $\overline{t_+}$ and the other half occupation 0, generating vast fluctuations on the overall statistics which are caused by the constrained binary structure of the graph.

Concerning the entropy, performing the steepest descent approximation on \eqref{eq_grandca} to first order as in \eqref{eq_entropy_can} we have,
\begin{equation}\nonumber
\begin{split}
S_{BG}\simeq -T\ln \rho  - \sum \limits_{ij} \hat{p}_{ij} \sum_{q'} \lambda_{q'} \hat{c}_{q'}^{(ij)} + \ln(T!) + \sum\limits_{ij} \ln \Delta_{ij}.
\end{split}
\end{equation}
Using $e^{\sum\limits_{q'} \lambda_{q'} \hat{c}_{q'}^{(ij)}} (e^\rho -1)=\frac{\hat{p}_{ij}}{1-\hat{p}_{ij}}$,  \eqref{eq_grandca_occ} one is lead to
\begin{widetext}
\begin{equation}
\label{eq_grandca_entropy}
\begin{split}
S_{BG} &= - \left \{ \sum_{ij} \left (\hat{p}_{ij} \ln \hat{p}_{ij} + (1-\hat{p}_{ij})\ln(1-\hat{p}_{ij})\right)  \right\}  -\left \{ T \ln \rho - \ln T! -\ln(e^\rho-1) \langle E \rangle   \right \} \\
S_{BG} &= S_{bin} + S_{dist}.
\end{split}
\end{equation}
\end{widetext}
This expression has two clear contributions: The first term $S_{bin}$ is the entropy corresponding to the binary constrained topology \cite{Annibale2009, Bianconi2008, Bianconi2008b} while the second one $S_{dist}$ corresponds to the additional multi-edge distinguishable structure. In other words, $S_{bin}$ counts all the possible ways to select $E$ states out of a total number $N(N-1)$ while $S_{dist}$ refers to the possible ways to allocate the $T$ events on these $E$ \textit{surviving} states.

The second term in \eqref{eq_grandca_entropy} can be explicitly evaluated in two limiting cases: The \textit{dense} case which corresponds to the thermodynamic limit and the \textit{sparse} for which the binary and weighted structure are equal $\overline{t_+}=T/E  \to 1$ (hence $\langle E \rangle=E$ is fixed).

Considering the \textit{sparse} case, one has from \eqref{eq_lambert} $\overline{t_+} \to 1$ so $\rho \to 0$ and hence,
\begin{equation}
\label{eq_entropy_sparse}
\begin{split}
\lim_{\rho \to 0, T\to \langle E \rangle} S_{dist} = \ln E!
\end{split}
\end{equation}
which corresponds to the micro-canonical counting of configurations coming from valid permutations of distinguishable multi-edges over the fixed binary structure of $E$ \textit{surviving} occupied states.

The \textit{dense} case corresponds to $T\to \infty$, which implies from \eqref{eq_lambert} $\rho \to \overline{t_+} = T/\langle E \rangle$ and $\ln(T!) \simeq T\ln T -T$,
\begin{equation}
\label{eq_entropy_dense}
\lim_{T \to \infty} \frac{S_{dist}}{T} = \ln \langle E \rangle
\end{equation}
which has a Shannon form if we consider $p\equiv \langle E \rangle ^{-1}$ which would be the probability associated to a multinomial process of sorting $T$ events over the surviving $\langle E \rangle$ binary links with identical probability $p$. In this limit, the  difference between sorting $\langle E \rangle $ independent Poisson processes with mean $T/\langle E \rangle$ and sorting $\langle E \rangle$ Poisson processes excluding the zero-occupation events is negligible. 

In the following, we present some examples to clarify the usage of this new methodology.

\subsection{Fixed $E$,$T$}
\label{fix_ET}

We start by considering the most simple case where we only fix the total number of events $T=\sum\limits_{ij} t_{ij}$ and the total number of existing binary links $E=\sum\limits_{ij} \Theta(t_{ij})$ on the network, in analogy to the paradigmatic model of the Erdos-Renyi graph in binary networks \cite{Erdos60onthe}.

In this case we introduce two Lagrange multipliers ($\lambda, \theta$) to fix ($E,T$). Proceeding from equation \eqref{eq_grandca}, we readily obtain the saddle point equations,
\begin{equation}
\begin{split}
E&=\sum\limits_{ij} \langle \Theta(t_{ij}) \rangle = N(N-1)\frac{\chi (e^{\rho} -1 )}{\chi (e^\rho -1)+1} \\
T&=\sum\limits_{ij} \langle t_{ij} \rangle = N(N-1)\frac{\chi e^{\rho} \rho}{\chi (e^\rho -1)+1}.
\end{split}
\end{equation}
Where we identify $\chi =e^\lambda, \rho=e^\theta$. We see that the average occupation numbers and edge existence probability are constant and their average values proportional as expected. Using \eqref{eq_grandca_occ} we compute the relevant magnitudes,
\begin{equation}\label{eq_E_T}
\begin{split}
&\langle \Theta(t_{ij}) \rangle \equiv \hat{p} = \frac{E}{N(N-1)} = Cnt\\
&\langle t_{ij} \rangle = \frac{e^\rho \rho }{e^\rho-1} \hat{p} = \overline{t_+} \hat{p} = \frac{T}{E}\hat{p} = \frac{T}{N(N-1)} = Cnt\\
&\sigma^2_{t} = \frac{T}{E}\hat{p} (1+\rho - \frac{T}{E}\hat{p}) \quad \quad  \sigma^2_{\Theta} = \hat{p} \left (1 - \hat{p}\right).
\end{split}
\end{equation}

We recover the binary structure of the well-known Erdös-Renyi graph \cite{Erdos60onthe} as a result of the binary projection of a non-trivial multi-edge structure, which on average values fulfils $\langle t \rangle = \frac{T}{E} p = \overline{t_+} p$.

Despite the average occupation numbers over the ensemble being equal to the cases in section \ref{fixT_cano}, the underlying statistic is not. All the nodes (and states) in this case are statistically equivalent and their associated strengths and degrees (incoming and outgoing)  are proportional on average (since they are fluctuating quantities not being fixed by the constraints)
\begin{equation}
\begin{split}
\langle s \rangle &= \bar{s} = \frac{T}{E} \langle k \rangle = \frac{T}{E} \bar{k} = \frac{T}{E} \frac{E}{N}= \frac{T}{N}\\
\sqrt{\frac{\sigma^2_k}{\langle k \rangle^2}} &= \sqrt{\frac{1-\sum\limits_j \hat{p}^2_{ij}}{\bar{k}}}\\
\sqrt{\frac{\sigma^2_s}{\langle s \rangle^2}} &= \sqrt{\frac{1}{\overline{t_+} k} + \frac{\sum\limits_j \hat{p}(1-\hat{p})}{\bar{k}^2}} \to \sqrt{\frac{\sigma^2_k}{\langle k \rangle^2}} \quad \text{as } T\to \infty.
\end{split}
\end{equation}


The entropy is readily computed from \eqref{eq_grandca_entropy} yielding,
\begin{widetext}
\begin{equation}
\begin{split} \label{eq_S_ET}
&S_{BG} = S_{E-R} + S_{dist}\quad \quad  \quad \quad  S_{E-R}= - N(N-1) (\hat{p} \ln \hat{p} + (1-\hat{p}) \ln (1-\hat{p})) \\
&S_{dist} = \ln(T!) - T \ln \left(\overline {t_+} + W (-\overline{t_+} e^{-\overline{t_+}})\right) + E \ln \left ( e^{\overline {t_+} + W (-\overline{t_+} e^{-\overline{t_+}})} -1 \right )
\end{split}
\end{equation}
\end{widetext}
Expression \ref{eq_S_ET} can also be computed using combinatorial arguments: Consider a process in which one selects $E$ states out of $N(N-1)$ and then populates each state with a single event chosen out of a set of $T$ distinguishable entities, finally, the rest of the $T-E$ events are sorted in the $E$ \textit{surviving} states chosen in the first place. The counting of microstates reads,
\begin{equation}\label{eq_S_micro_ET}
\Gamma (E,T,N) = \binom{N(N-1)}{E}\frac{T!}{(T-E)!} E^{T-E}
\end{equation}
and hence the micro-canonical entropy is $S_{BG} = \ln \Gamma(E,T,N)$. Here again one recovers the equivalence with \eqref{eq_S_ET} in both the \textit{sparse} and \textit{dense} limits considered earlier. 

\subsection{Fixed degree sequence $\vec k=\{(k^{out},k^{in})_i\}\,, T$}\label{fixK}

The next important case to consider is the one where the node binary connectivity of the graph is fixed. Such a situation is specially interesting as our framework permits to understand binary networks as the projection (for instance due to partial information or limited resolution) of a process generated by independent agents.

In this case the constraints read,
\begin{equation}\label{eq_cons_kk}
\begin{split}
\hat{C}_i(t_{ij})&= \sum_i \Theta(t_{ij}) = k_i^{out}\\ \hat{C}_j(t_{ij})&= \sum_j \Theta(t_{ij}) = k_j^{in}.
\end{split}
\end{equation}
So we introduce two sets of Lagrange multipliers ($\{\lambda_i,\beta_i\}$) and for expression \eqref{eq_grandca} we have,
\begin{widetext}
\begin{equation}\nonumber
\begin{split}
\Omega&= \int d\vec \alpha d\vec \beta \exp \left \{-\sum_i \alpha_i k_i^{out} -\sum_j \beta_j k_j^{in} \right \}
 \exp \left \{ - \theta T + \ln T! +  \sum\limits_{ij} \ln \left [ e^{\alpha_i+\beta_j+b_{ij}} \left(e^{e^{h_{ij}+\theta}}-1 \right) +1 \right]\right \}=\\
&= \int d\vec \alpha d\vec\beta \exp g(\{\alpha_i,\beta_i,h_{ij},b_{ij}\},\theta).
\end{split}
\end{equation}
\end{widetext}
And we solve the saddle point equations,

\begin{equation}\label{conf1}
\begin{split}
\left. \partial_\theta g\right|_{h_{ij}=b_{ij}=0 \, \forall ij} &=0 \implies T=\sum_{i,j} \frac{x_i y_j e^\rho \rho}{x_iy_j\left ( e^{\rho}-1 \right)+1}
\\
\left.\partial_{\alpha_i} g\right|_{h_{ij}=b_{ij}=0 \, \forall ij}&= 0 \implies k_i^{out} = \sum_j \frac{x_iy_j\left ( e^{\rho}-1 \right)}{x_iy_j\left ( e^{\rho} -1 \right)+1}
\\
\left.\partial_{\beta_j} g\right|_{h_{ij}=b_{ij}=0 \, \forall ij} &= 0 \implies k_j^{in} = \sum_i \frac{x_iy_j\left ( e^{\rho}-1 \right)}{x_iy_j\left ( e^{\rho}  -1 \right)+1},
\end{split}
\end{equation}
where we have identified $x_i\equiv e^{\alpha_i}$, $y_j\equiv e^{\beta_j}$ and $\rho \equiv e^\theta$. The occupation numbers and binary occupation probability respectively read,
\begin{equation}\label{eq_fix_k_occ}
\begin{split}
\langle t_{ij} \rangle &=\frac{x_i y_j e^{\rho} \rho}{x_iy_j\left ( e^{\rho}  -1 \right)+1} \\
\hat{p}_{ij} &= \frac{x_i y_j (e^{\rho}-1)}{x_iy_j\left ( e^{\rho}-1 \right)+1}.
\end{split}
\end{equation}

As expected we recover the proportionality of strengths and degrees \cite{Barrat2004a} for each node under this particular set of constraints.
\begin{equation}
\langle t_{ij} \rangle = \hat{p}_{ij} \frac{e^\rho \rho}{e^{\rho}-1} \implies \langle s_i \rangle = \bar{t}  k_i = \frac{T}{E} k_i.
\end{equation}

Considering only the binary projection of the graph, one gets,
\begin{equation}
\hat{p}_{ij} = \frac{\mu \kappa_i \lambda_j}{\mu \kappa_i \lambda_j +1 }.
\end{equation}
Where we have identified $\kappa_i \equiv x_i$, $\lambda_j \equiv y_j$ and $\mu \equiv (e^{\rho}-1) $.

The previous expression corresponds exactly with the expression for the so called \textit{canonical ensemble} of the random graph with any given degree distribution \cite{Boguna2003},\cite{Park2003}, where for fixed $N$, the $\mu$ parameter controls the edge density \cite{Anand2011}. And since $x_i$ is a quantity related with node $i$ (hence related with $k_i$), we obtain again the structural correlations of the configuration graph model \cite{PhysRevE.67.046118}.

\section{Multi-edge network with given linear constraints depending on both the the occupation numbers $t_{ij}$ and their binary projection $\Theta(t_{ij})$}
\label{seq_kss}
We analyze for the sake of completeness the most general case where both types of considered constraints are fixed, \eqref{eq_cons_cann} and  \eqref{eq_binary_cons}.

We introduce two sets of Lagrangian multipliers for the multi-edge constraints $\{\theta_q\}$ and the binary ones $\{\lambda_{q'}\}$, plus an additional $\theta$ corresponding to the constraint on the total number of events. The procedure then is analogous to the one in the previous section yielding finally from \eqref{eq_grandca},
\begin{widetext}
\begin{equation}\label{eq_grandca_max}
\begin{split}
&\Omega= \int d\vec \theta e^{-\theta T} e^{-\sum_{q'} \lambda_{q'} \hat{C}_{q'}} e^{-\sum_q \theta_q C_q} \prod_{q} d\vec \theta_q d\vec \lambda_{q'} T!
\exp \left \{ \sum \limits_{ij} \ln \left \{ e^{b_{ij}+ \sum \limits_{q'} \lambda_{q'} \hat{c}_{q'}^{(ij)}} \left (e^{e^{h_{ij}+\theta+\sum_q \theta_q c_q^{(ij)}}} -1\right)+1 \right \} \right\}.
\end{split}
\end{equation}
\end{widetext}
Using \eqref{eq_der_cum}, \eqref{eq_der_cum_bin} we obtain the statistic for the occupation numbers and their projections,

\begin{equation}\label{eq_fix_bin_nonbin}
\begin{split}
\langle t_{ij} \rangle =& \frac{1}{\Delta_{ij}} e^{\sum_{q'} \lambda_{q'} \hat{c}_{q'}^{(ij)}} \exp \left (e^{\theta + \sum_q \theta_q c_q^{(ij)}} \right) e^{\theta+\sum_q \theta_qc_q^{(ij)}}\\
\sigma_{t_{ij}} =& \langle t_{ij} \rangle \left (1+ e^{\theta + \sum_q \theta_q c_q^{(ij)}} - \langle t_{ij} \rangle \right)\\
\hat{p}_{ij} =& \frac{1}{\Delta_{ij}} e^{\sum_{q'} \lambda_{q'} \hat{c}_{q'}^{(ij)} } \left( e^{e^{\theta + \sum_q \theta_q c_q^{(ij)}}}\right)\\
\sigma_{\Theta(t_{ij})} =& \hat{p}_{ij}(1-\hat{p}_{ij})\\
\Delta_{ij} =& e^{\sum_{q'} \lambda_{q'} \hat{c}^{(ij)}_{q'}} \left( e^{e^{\theta + \sum_q \theta_q c_q^{(ij)}}} -1 \right) +1.
\end{split}
\end{equation}

Assuming that the saddle point equations can be solved, which means that the imposed combination of binary and multi-edge constraints is graphical, i.e. $s_i \geq k_i\, \forall i \in [1,N]$ for the case of fixed strength and degree sequence for instance, then the probability to obtain a graph can still be written in terms of the Lagrange multipliers as a sum of independent ZIP processes with different parameters,
\begin{equation}\label{eq_grandca_gen}
\begin{split}
&P(\mathbf{T}|\{\hat{p}_{ij},\mu_{ij}\} ) = \prod_{i,j}(1- \hat{p}_{ij}) )^{1-\Theta(t_{ij})} \left \{ \frac{\hat{p}_{ij}}{e^{\mu_{ij}} -1} \frac{\mu_{ij}^{t_{ij}}}{t_{ij}!} \right \}^{\Theta(t_{ij})}
\end{split}
\end{equation}
where $\mu_{ij} = e^{\theta + \sum_q \theta_q c^{(ij)}_q}$, is a quantity related to the average value of the occupation number of the given link, conditioned that this link exists,
\begin{equation}\label{eq_pos_tij}
\frac{\sum\limits_{t=1}^{\infty} t_{ij} P(t_{ij}|\mu_{ij},\hat{p}_{ij},t_{ij}>0)}{\sum\limits_{t=1}^{\infty} P(t_{ij}|\mu_{ij},\hat{p}_{ij},t_{ij}>0)} = \mu_{ij} \frac{e^{\mu_{ij}}}{e^{\mu_{ij}} -1} =  \langle t_{ij}^+ \rangle.
\end{equation}
The relative fluctuations on occupation numbers do not vanish in the thermodynamic limit due to the strong constraints imposed by the binary structure. One can still express $\mu_{ij} (\langle t^+_{ij} \rangle)$ using \eqref{eq_lambert},
\begin{equation}
\mu_{ij} (\langle t^+_{ij} \rangle) = \langle t^+_{ij} \rangle + W(-\langle t^+_{ij} \rangle e^{-\langle t^+_{ij} \rangle}).
\end{equation}
In the thermodynamic limit ($T\to \infty$ which implies $\langle t^+_{ij} \rangle \to \infty$), expression \eqref{eq_pos_tij} converges to $\mu_{ij} \simeq \langle t_{ij}^+ \rangle$.

Regarding the relation between expected occupation numbers and their binary projections, one has $\langle t_{ij} \rangle = \langle t_{ij}^{+} \rangle \hat{p}_{ij}$ and the constant relation of proportionality is broken $\frac{\langle t_{ij} \rangle}{\hat{p}_{ij}} \neq  Cnt $. This extends the well known result that it is impossible to generate uncorrelated networks both at the level of strengths and degrees for multi-edge networks or weighted networks \cite{Garlaschelli2009}, \cite{Serrano2006}.

Concerning the entropy, approximating \eqref{eq_grandca} by saddle point methods using \eqref{eq_fix_bin_nonbin} we obtain the general expression that includes all the previous cases considered,
\begin{equation}\label{eq_grandcaCA_entropy}
\begin{split}
S_{BG} &= S_{dist} + S_{bin} \\
S_{dist} & = \ln T! + \sum\limits_{ij} \hat{p}_{ij} \ln \left(e^{\mu_{ij}} -1 \right) - \sum \limits_{ij} \langle t_{ij} \rangle \ln \mu_{ij}.
\end{split}
\end{equation}
where $S_{bin}$ is still the binary contribution to the entropy (with the same form as in \eqref{eq_grandca_entropy}).

The two limiting cases early considered can again be evaluated. The \textit{sparse} case implies that $\overline{t_{ij}^+} = \frac{T}{E} \to 1 \, \forall i,j$ which means $\mu_{ij} \to 0$ and $\langle t_{ij} \rangle \to \hat{p}_{ij}$ obtainig,
\begin{equation}
\lim_{T\to E} S_{dist} =\ln E!
\end{equation}
which is identical to the previous one (since it is equivalent to dropping the strength constraints).

For the thermodynamic limit (\textit{dense} case), we have $e^{\mu_{ij}}-1 \to e^{\mu_{ij}} \to e^{\langle t^+_{ij}\rangle }$ and then,
\begin{equation}
\lim_{T\to \infty} \frac{S_{dist}}{T} = -  \sum\limits_{ij} \frac{\langle t_{ij} \rangle}{T} \ln \frac{\langle t^+_{ij} \rangle}{T}= -  \sum\limits_{ij} \frac{ \langle t_{ij} \rangle}{T} \ln \frac{\langle t_{ij}\rangle}{T \hat{p}_{ij}}.
\end{equation}
for which the previous expressions encountered are limiting cases. On one hand, if we relax the constraints on the occupation numbers, then $\langle t_{ij}^+\rangle = \overline{t_+} = T/E \, \forall i,j$ and we recover expression \eqref{eq_entropy_dense}. On the other hand, not fixing any binary related quantity implies that as $T\to \infty$, $\hat{p}_{ij} = 1- e^{-\langle t_{ij} \rangle} \to 1$ (fully connected topology) and we are lead to \eqref{eq_entro_can}. Finally, not fixing any constraints, $\langle t^+_{ij} \rangle = \langle t_{ij} \rangle = \bar{t}$, we recover \eqref{eq_nocons_entr}.

For simplicity, the only example we report in this section corresponds to the very relevant case in which both strength and degree sequences are fixed, the rest of cases being easily derivable from the general theory exposed.

\subsection{Fixed relative strength sequence and fixed degree sequence $\vec s=\{(s^{out},s^{in})_i\},\, \vec k =\{(k^{out},k^{in})_i\}$}
\label{fixKS}

We analyze here a situation where the constraints imposed are the strength and degree sequence (\eqref{eq_cons_ss} and \eqref{eq_cons_kk}). The calculations are analogous to the previous section obtaining,

\begin{equation}\nonumber
\begin{split}
\left.\partial_\theta h\right|_{h_{ij}=b_{ij}=0\,\forall ij} &= 0 \implies T=\sum\limits_{ij} \frac{1}{\Delta_{ij}} x_iy_j e^{z_i w_j e^\theta} e^{\theta} z_i w_j \\ 
\left.\partial_{\alpha_i} h\right|_{h_{ij}=b_{ij}=0\,\forall ij} &= 0 \implies k_i^{out} = \sum_j \frac{1}{\Delta_{ij}}x_iy_j \left( e^{z_iw_je^\theta} -1\right)\\
\left.\partial_{\beta_j} h\right|_{h_{ij}=b_{ij}=0\,\forall ij} &= 0 \implies k_j^{in} = \sum_i \frac{1}{\Delta_{ij}}x_iy_j \left( e^{z_iw_je^\theta} -1\right)\\
\left.\partial_{\gamma_i} h\right|_{h_{ij}=b_{ij}=0\,\forall ij} &= 0 \implies s_i^{out} = \sum_j
\frac{1}{\Delta_{ij}}x_iy_j z_i w_j e^{z_iw_je^\theta}e^\theta  \\
\left.\partial_{\epsilon_j} h\right|_{h_{ij}=b_{ij}=0\,\forall ij} &= 0 \implies s_j^{in} = \sum_i
\frac{1}{\Delta_{ij}}x_iy_j z_i w_j e^{z_iw_je^\theta}e^\theta\\
\Delta_{ij} & \equiv x_iy_j \left\{ \exp \left (e^\theta z_i w_j \right )-1 \right\} +1
\end{split}
\end{equation}
where we have identified $x_i\equiv e^{\lambda^{(k_{out})}_i},y_j\equiv e^{\lambda^{(k_{in})}_j},z_i\equiv e^{\theta^{(s_{out})}_i},w_j\equiv e^{\theta^{(s_{in})}_j}$ corresponding to the $4N$ Lagrange multipliers introduced. We hence obtain the saddle point equations,
\begin{equation}\label{eq_fixKS}
\begin{split}
\langle t_{ij} \rangle =& \frac{1}{\Delta_{ij}} x_iy_j \exp \left ( e^\theta z_i w_j \right) z_iw_je^\theta\\
\sigma_{t_{ij}} =& \langle t_{ij} \rangle \left (z_i w_j e^{\theta} +1 - \langle t_{ij} \rangle \right)\\
\hat{p}_{ij} =& \frac{1}{\Delta_{ij}} x_iy_j \left( e^{z_iw_je^\theta}-1\right)\\ \sigma_{\Theta(t_{ij})} =& \hat{p}_{ij}(1-\hat{p}_{ij}).
\end{split}
\end{equation}

Note that the expressions found in the previous cases are particular examples of this general problem and can be readily recovered by removing the appropriate constraints, i.e., making the Lagrange multipliers equal to zero, which in this case is equivalent to setting $x_i=y_j=1\, \forall i,j$ or $z_i=w_j=1\, \forall i,j$ or both. 

We can revisit the case where only the strength sequence is fixed. Now, although the resulting statistics are Poisson and not multinomial,  it can be proved that in the thermodynamic limit both descriptions are equivalent (see appendix \ref{ap_equiv}). Additionally, in such a case one can obtain the statistics of the binary projection of the occupation numbers $\hat{p}_{ij} = 1 - e^{-\langle t_{ij} \rangle}$.


Unfortunately, the explicit form of the Lagrange multipliers for the degrees or the strengths cannot be solved, since the uncorrelated approximation is no longer valid,
\begin{equation}
\begin{split}
&x_i y_j (e^{e^\theta z_i w_j} -1 ) \simeq e^{e^\theta z_i w_j} x_i y_j\\
&\text{if }x_i y_j e^{e^\theta z_i w_j} << 1 \implies \hat{p}_{ij} \simeq x_iy_j e^{e^\theta z_i w_j}.
\end{split}
\end{equation}
In this last expression the factorization of the connection probability in two node-dependent magnitudes is impossible, despite the approximation \textit{assumed}. Hence one sees again that there is no way of generating uncorrelated networks at the level of degrees under the strict set of constraints considered.


\section{Conclusions}
\label{sec_concl}

The present work deals with the statistical framework of the so-called \textit{weighted networks}, already studied in \cite{Park2004,Garlaschelli2009} for the case of indistinguishable entities and completed here for the case of distinguishable units. The decision upon which model to take depends on the kind of (physical) process is generating the network at study. We have started by properly defining the differences between weighted and multi-edge networks based on the distinguishability or not of the elements forming a network. We have then properly set up a framework of multi-edge networks in a statistical mechanics approach by defining appropriate thermodynamic limits which can be mapped to a system of classical particles populating a finite set of discrete levels.

We have obtained analytical expressions for general cases with constraints depending linearly on the occupation numbers as well as their binary projections and some common interesting cases have been developed. Previous results found in the literature have been recovered, specifically the correlations of the configurational model (for binary constraints on the degree sequence) and the absence of correlation between occupation numbers and degree once the degree sequence is fixed among others. Our treatment uncovers explicit relations between the binary occupation probability of an edge and its expected occupation number. These results permit an extensive treatment of the finite size effects present in this kind of networks, since they are fully valid both in the thermodynamic limit and intermediate cases.

Furthermore, we have presented general forms for the probability of obtaining a graph with given constraints which can be also easily extended to the canonical and grand-canonical ensembles. As a complement we have also introduced the main ideas which can lead to the efficient generation of multi-edge graphs though its details are left for development in future work.

The applications of the theory developed can be extended to a wide variety of fields, specially in the very active transportation research area and human mobility subjects \cite{Barthelemy2011}, which have received a renewed interest in the last times. It also opens the door to a proper multiplex extensions of this kind of networks and the analysis of similar systems in terms of entropy measures \cite{Bianconi2009}.


\acknowledgments This work has been partially supported by the
Spanish DGICYT Grant FIS2009-13364-C02-01 and by the Generalitat de Catalunya 2009-SGR-00838. 
O.S. have been supported by the Generalitat de Catalunya through the FI Program.

\appendix
\section{Cumulant generating functions}
\label{ap_cumul}
In this section we present the cumulant generating functions for the different models proposed (Bernoulli, Multinomial, Poisson and Zero Inflated Poisson) and show that their close relationship to the expressions developed in the main text.

The cumulant generating function of the probability distribution $P(h)$ of a variable $h$ is defined as $K(h,x) = \ln M(h,x)$, being $M(h,x)=\langle e^{hx} \rangle$ its moment generating function and $x$ and auxiliary field. Once $K(h,x)$ is known, all central cumulants $\kappa_k$ can be obtained by derivation, uniquely determining the distribution.
\begin{equation}
\begin{split}
\kappa_k = \left . \partial_k K(h,x)\right |_{x=0}.
\end{split}
\end{equation}
Note that if we consider the joint distribution of two (or more) independent variables $h_1, h_2$, being it a product of the individual distributions $P(h_1),P(h_2)$, then $M_{1,2}(h_1,h_2,x_1,x_2) = M_1(h_1,x_1) M_2 (h_2,x_2)$ and finally its joint cumulant generating function factorizes in the sum $K_{12}(h_1,h_2,x_1,x_2) = K_1(h_1,x_1) + K_2(h_2,x_2)$.

Having introduced that, if we start at the micro-canonical level (equation \eqref{eq7}) and identify,
\begin{widetext}
\begin{equation}
\begin{split}
&P(\{ t_{ij} \}) = \Omega^{-1}\prod_q \delta(C_q - C_q(\{t_{ij}\}) = \left \{ \sum \limits_{\{c\}} \prod\limits_{q}^Q \delta(C_q - C_q(\{t_{ij}\})\right \}^{-1} \prod_q \delta(C_q - C_q(\{t_{ij}\}) \equiv  C^{-1} \prod_q \delta(C_q - C_q(\{t_{ij}\})\\
&\ln \Omega(\{h_{ij}\}) = \ln \left\{ C \sum_{\{c\}} P(\{t_{ij}\}) e^{h_{ij}}\right\} =\ln C + K(\{t_{ij}\},\{h_{ij}\}),
\end{split}
\end{equation}
\end{widetext}
we clearly see that,
\begin{equation}
\partial_{h_{ij}} \ln \Omega (\{t_{ij}\},\{h_{ij}\}) = \partial_{h_{ij}} K(\{t_{ij}\},\{h_{ij}\})
\end{equation}
and the relation between both objects is apparent.

Starting at the level where only the number of events $T$ is fixed (where all the distributions reduce to a Multinomial form with associated probabilities $\{p_{ij}\}$), we take equation \eqref{eq_cumul_multi} and perform the saddle point approximation to obtain,
\begin{equation}
\begin{split}
\ln \Omega &\simeq T \ln \sum\limits_{ij} \exp \left((h_{ij}+\sum_q\theta_q c^{(ij)}_q) \right) + F(\{\theta_q\}, \{C_q\}) =\\
&= H(\{h_{ij}\}) + F(\{\theta_q\}, \{C_q\}).
\end{split}
\end{equation}

Despite having additional terms, with regards to differentiation with respect to $h_{ij}$ (needed to recover the moments of the distribution of $t_{ij}$), the form obtained is always of the type $H(\{h_{ij}\}) =  T \ln \sum \pi_{ij} e^{h_{ij}}$, (since $F$ is a function not depending on the auxiliary fields $\{h_{ij}\}$) and hence the underlying statistic is multinomial because $\sum\limits_{ij}\pi_{ij} = 1$.

Considering now only the most general case in equation \eqref{eq_grandca} we have again (adding external fields for the binary projection $\{b_{ij}\}$),

\begin{equation}
\begin{split}
\ln \Omega &\simeq \sum \limits_{ij} \ln \left \{ e^{b_{ij}+\sum\limits_{q'} \lambda_{q'} \hat{c}_{q'}^{(ij)}} \left (e^{e^{h_{ij}+\sum\limits_{q} \theta_q c^{(ij)_q}}} -1\right)+1 \right \} \\
&+ F(\{\theta_q\},\{\lambda_{q'}\},\{C_q\},\{C_{q'}\})
\end{split}
\end{equation}
Dropping the constraints on the binary projection, i.e. $\lambda_{q'} = b_{ij}=0 \forall q,ij$ yields,
\begin{equation}
H(\{h_{ij}\}) = \sum_{i,j} e^{h_{ij}} \mu_{ij},
\end{equation}
where we identified $\mu_{ij} = e^{ \sum\limits_q \theta_q c_q^{(ij)}}$. This expression (up to derivation with respect to $h_{ij}$) has the same form as a sum of independent Poisson cumulant generating functions ($\mu e^{h_{ij}}$).

Dropping the constraints on the multi-link nature of the network $\theta_q = h_{ij}=0 \forall q,ij$ (except the one on the total number of multi-links, $\theta$), we have,
\begin{equation}
H(\{q_{ij}\}) = \sum_{i,j} \left \{\ln \left (\hat{p}_{ij} e^{b_{ij}} +  (1-\hat{p}_{ij}) \right ) - \ln \left ( 1- \hat{p}_{ij} \right ) \right\},
\end{equation}
where we identified $\hat{p}_{ij}$ from equation \eqref{eq_fix_k_occ}. And we see that the prior expression is closely related to the cumulant generating function of independent Bernoulli processes (with respect again to derivation on $\{q_{ij}\}$ terms).

Finally, the most general case can be mapped to a mixed Zero Inflated Poisson process as we shall prove: Imagine the outcome of a process in which we sort $N(N-1)$ independent Bernoulli processes and from the result of it, if the outcome is positive, we sort a Poisson process on top of it (discarding the no-occurrence event). Since the processes are independent, we shall consider a single one of them and then write the overall probability of the events as the product of the different probabilities $P(t)$. The associated probability of the event just described is,
\begin{equation}
P_{b-p}(t) = (1-p)^{\Theta(t)} \left ( \frac{p}{e^{\mu}-1} \frac{\mu^t}{t!}\right)^{\Theta(t)}.
\end{equation}
Which represents a probability measure over an integer quantity. We can compute the mean and variance of $t$ yielding,
\begin{equation}
\begin{split}
\langle t \rangle &= 0 + \frac{p}{e^\mu-1} \sum\limits_{t=1}^\infty t \frac{\mu^t}{t!} = p\frac{\mu e^\mu}{e^\mu-1}\\
\sigma^2_{t} &= \langle t \rangle (1+\mu - \langle t \rangle)\\
\langle \Theta(t) \rangle &= p; \quad \quad \sigma^2_{\Theta(t)} = p(1-p).
\end{split}
\end{equation}
The obtained expressions need to be compared with \eqref{eq_fixKS}, which allows to identify $\mu = e^\theta \prod_q e^{\theta_q c_q^{(ij)}}$ and $p=\hat{p}_{ij}$. Moreover, concerning the cumulant generating function, one finds,
\begin{widetext}
\begin{equation}
\begin{split}
\ln \langle e^{ht} \rangle &= \ln \sum\limits_{t=0}^\infty P_{b-p}(t) e^{ht}= \ln \left \{ 1-p + \frac{p}{e^\mu-1} \left ( e^{\mu e^h} -1  \right) \right \} = \ln(1+p) +  \ln \left \{1+\frac{p}{1-p}\frac{\left ( e^{\mu e^h}-1 \right)}{e^\mu-1}   \right \},
\end{split}
\end{equation}
\end{widetext}
which is identical to the argument in the sum of equation \eqref{eq_grandca} (except for a linear constant) and captures the more general case considered.

\section{Ensemble equivalence and graph generation}
\label{ap_equiv}
Throughout this paper we have uncovered the mathematical expressions allowing to generate networks under different ensembles using a probabilistic framework over $t_{ij}$. Explicitly they can be summarized,
\begin{itemize}
\item Canonical Ensemble (linear constraints on $t_{ij}$):
\begin{equation}
P(\mathbf{T}|T,\{\theta_q\}) = \frac{T!}{\prod\limits_{ij} t_{ij}!} \prod \limits_{ij} p_{ij}^{t_{ij}}
\end{equation}
\item Grandcanonical Ensemble (linear constraints on $\Theta(t_{ij})$ and/or on $t_{ij}$):
\begin{equation}\label{eq_grand_gen}
P(\mathbf{T}|\{\hat{p}_{ij},\mu_{ij}\} ) = \prod_{i,j}(1- \hat{p}_{ij}) )^{1-\Theta(t_{ij})} \left \{ \frac{\hat{p}_{ij}}{e^{\mu_{ij}}-1} \frac{\mu_{ij}^{t_{ij}}}{t_{ij}!} \right \}^{\Theta(t_{ij})}
\end{equation}
\item Microcanonical Ensemble: This ensemble can be used by generating sequences of $\{t_{ij}\}$ using the two above expressions and discarding those not corresponding exactly with the imposed constraints.
\end{itemize}

We have shown already that the relative fluctuations of the linear constraints on the occupation numbers $t_{ij}$ vanish in the thermodynamic limit and that the binary depending constraints are non-vanishing in this limit. We finally prove here that the grand-canonical ensemble considering only linear constraints on occupation numbers is strictly equivalent to the canonical and the microcanonical in the thermodynamic limit.

To do so, we make use of the properties of the multinomial distribution to recover it under the cases where no constraints or only linear constraints on $t_{ij}$ are imposed. In this case $p_{ij}=1-e^{-\mu_{ij}}$ \eqref{eq_fix_bin_nonbin} and \eqref{eq_grand_gen} reduce to the product of Poisson distributions with different mean parameters $\mu_{ij}$.

The outcome of a process in which we sort different independent Poisson variables of parameters $\{\mu_{ij}\}$ can be equivalently expressed as a product of a multinomial process of $\langle T \rangle = \sum \mu_{ij}$ multinomial trials with associated probabilities $\{p_{ij} = \frac{\mu_{ij}}{\langle T \rangle}\}$. Hence we have that,
\begin{equation}\label{multipois}
P(\mathbf{T}) = \text{Mult}(\{ p_{ij}\} | T) \text{Pois}(\lambda_T = \sum \mu_{ij}).
\end{equation}

And since the resulting occupation numbers statistics derive to a Poisson distribution, which has vanishing fluctuations on the thermodynamic limit ($\sigma^2_{t_{ij}} / \mu_{ij}^2 = \mu_{ij}^{-1} \propto T^{-1}$ , then the equivalence between the two presented calculations is completely proved.
\section{finite T in eq.\ref{eq_grandca} }
\label{ap_finite}
\begin{figure}[hbp]
\begin{center}
\includegraphics[scale=0.45]{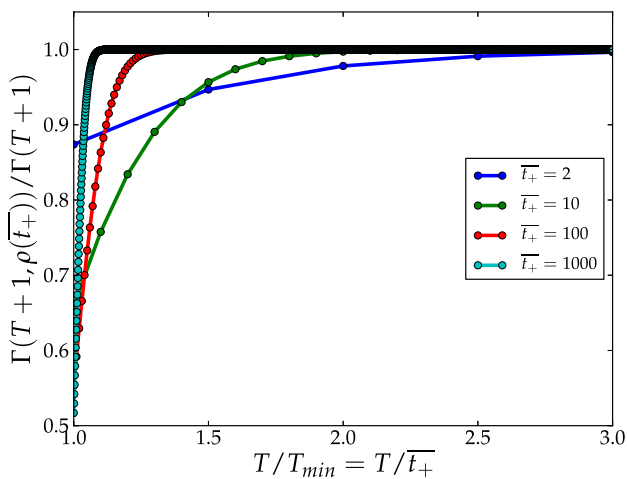}
\caption{$\frac{\sum \limits_1^T \frac{\rho^t}{t!}}{e^{\rho}} = \frac{\Gamma(T+1, \rho(\overline{t_{+}})}{\Gamma(T+1)}$ for different values of $\overline{t_+}$ and $T$, we observe the extreme rapid convergence to unity as $T$ grows (note that $T\geq \overline{t_+} = T/E$).}
\label{fig_ap1}
\end{center}
\end{figure}

Let us assume that instead of having $T \rightarrow \infty$ in \eqref{eq_grandca} we consider $T$ finite. Therefore
\begin{widetext}
\begin{equation}\label{eq_grandca_finite}
\begin{split}
\Omega&=\int e^{-\theta T} e^{-\sum_{q'} \lambda_{q'} C_{q'}} d\vec \theta \prod_{q'}  d\vec \lambda_{q'} T! \left [ e^{b_{ij}+\sum_{q'} \lambda_q' \hat{c}_{q'}^{(ij)}} \sum_{t_{ij}'=1}^T \frac{e^{(h_{ij}+\theta)t_{ij}'}}{t_{ij}'!} + \frac{1}{0!}\right ] \\
&\sum\limits_{\sum t_{ij} = T-t_{ij}'} \prod\frac{\left( e^{(b_{ij}+\sum_{q'}\lambda_{q'} \hat{c}_{q'}^{(ij)}) \Theta(t_{ij}) +h_{ij} + \theta}\right)^{t_{ij}}}{t_{ij}!} \\
\end{split},
\end{equation}
\end{widetext}
we now perform the finite sum inside the integral
\begin{equation}
\sum \limits_1^T \frac{z^t}{t!} = \frac{\Gamma(T+1, z)e^z}{\Gamma(T+1)}.
\end{equation}
And for the occupation numbers and edge probability obtain,
\begin{equation}
\begin{split}
\langle t_{ij} \rangle &= \frac{1}{\Delta_{ij}} e^{\sum_{q'} \lambda_{q'} \hat{c}_{q'}^{(ij)}} e^{\theta} \left (  e^{e^\theta} \frac{\Gamma(T+1, e^{\theta})}{\Gamma(T+1)}  - \frac{e^{T\theta}}{\Gamma(T+1)} \right )\\
\hat{p}_{ij} &= \frac{1}{\Delta_{ij}} e^{\sum_{q'} \lambda_{q'} \hat{c}_{q'}^{(ij)}} (e^{e^{\theta}} \frac{\Gamma(T+1, e^{\theta})}{\Gamma(T+1)}-1)\\
\Delta_{ij} &= e^{\sum_{q'} \lambda_{q'} \hat{c}_{q'}^{(ij)}} (e^{e^\theta} -1 ) +1 ),
\end{split}
\end{equation}
which converge extremely quickly to the obtained results for $T\to \infty$ (see figure \ref{fig_ap1}).

\section{Additional examples. Fixed binned distribution of costs $ \{c_n,N_{c_n}\}$}
\label{ap_extra}

We here report and additional example which may be of interest for studies on transportation origin-destination matrices, where some forms of trip-cost distribution have been discussed \cite{Bazzani2010}, \cite{Liang2012a}.

Starting from section \eqref{fixTC}, it is a matter of considering the additional term,
\begin{equation}\nonumber
\exp \left ( \sum\limits_n \kappa_n \xi_n(d_{ij}) \right)
\end{equation}
on the equations, where $\kappa_n$ are additional Lagrange multipliers satisfying that,
\begin{equation}\nonumber
N_n = \sum\limits_{ij} \kappa_n \xi_n(d_{ij}) t_{ij}
\end{equation}
Where $N_n$ is the number of trips whose distance is in the interval $[d_{n-1},d_n)$ and $\xi_n$ is the indicator function of such an event. The size of the bins needs to be chosen in an appropriately manner as to give consistency to the distribution obtained. Such an example is particularly important for its importance to assess whether an observed \textit{O-D} is caused by a particular tendency of the agents that create it to move or conversely, the space where they move shapes the form of the obtained \textit{O-D}.

The expressions of $p_{ij}$ in this case read,
\begin{equation}\nonumber
p_{ij}=\frac{e^{\sum_{n} \kappa_n \xi_(d_{ij})}}{\sum\limits_{ij} e^{\sum_{n} \kappa_n \xi_(d_{ij})}}.
\end{equation}
And the prior considerations are also valid. Note also that keeping the multinomial framework, all the quantities considered are intensive, while the expected strength and average occupation numbers remain extensive variables. Additionally, in the thermodynamic limit these quantities have vanishing relative fluctuations.



\bibliography{library_good}
\end{document}